\documentclass[aps,prb,twocolumn,showpacs,superscriptaddress]{revtex4-1}
\usepackage[latin1]{inputenc}
\usepackage{amsmath}
\usepackage{amssymb}
\usepackage{siunitx}
\usepackage[english]{babel}
\usepackage{graphicx, color}
\usepackage{hyperref}
\definecolor{link}{rgb}{0.1,0.1,0.9}
\hypersetup{colorlinks=true,linkcolor=link,citecolor=link,urlcolor=link,linktocpage}
\usepackage{epstopdf}
\usepackage{color}
\usepackage{longtable}
\usepackage{natbib}
\usepackage{mhchem}

\begin{document}

\title{Rare-earth tuned magnetism and magnetocaloric effects in double perovskites $R_2$NiMnO$_6$}

\author{Anzar Ali}
\affiliation{Department of Physical Sciences, Indian Institute of Science Education and Research Mohali, Knowledge city, Sector 81, SAS Nagar, Manauli PO 140306, Mohali, Punjab, India}

\author{Kanika Pasrija}
\affiliation{Department of Physics, Panjab University, Sector 14, PO 160014, Chandigarh,  India}

\author{Gyaneshwar Sharma}
\affiliation{Department of Physical Sciences, Indian Institute of Science Education and Research Mohali, Knowledge city, Sector 81, SAS Nagar, Manauli PO 140306, Mohali, Punjab, India}
\affiliation{Department of Physics, TDPG College Jaunpur, PO 222002, Uttar Pradesh, India}

\author{Sanjeev Kumar} \email{sanjeev@iisermohali.ac.in}
\affiliation{Department of Physical Sciences, Indian Institute of Science Education and Research Mohali, Knowledge city, Sector 81, SAS Nagar, Manauli PO 140306, Mohali, Punjab, India}

\author{Yogesh Singh} \email{yogesh@iisermohali.ac.in}
\affiliation{Department of Physical Sciences, Indian Institute of Science Education and Research Mohali, Knowledge city, Sector 81, SAS Nagar, Manauli PO 140306, Mohali, Punjab, India}

\begin{abstract}
\noindent
We present a comprehensive experimental study of magnetization and magnetocaloric effect (MCE) in double perovskite (DP) materials $R_2$NiMnO$_6$ with $R =$ Pr, Nd, Sm, Gd, Tb, and Dy. While a paramagnetic to ferromagnetic transition, with T$_{\rm C}$ in the range $\sim 100 - 200~$K, is a common feature that can be attributed to the ordering of Mn$^{4+}$ and Ni$^{2+}$ magnetic moments, qualitatively distinct behavior depending on the choice of $R$ is observed at low temperatures. These low-temperature anomalies in magnetization are also manifest in the change in magnetic entropy, $-\Delta S_{M}$, whose sign depends on the choice of $R$. In order to understand these results, we present theoretical analysis based on mean-field approximation and Monte Carlo simulations on a minimal spin model. The model correctly captures the key features of the experimental observations.  
\end{abstract}

\maketitle

\section{Introduction}
Magnetic materials have undoubtedly played an important role in advancing the technology to its commonly used current form  \cite{Gutfleisch_2011}. What underlies these technologies are the key concepts or phenomena in magnetism that have been harnessed to our advantage via a careful design. Some of the technologically useful phenomena displayed by magnetic materials are, giant magnetoresistance, magnetocapacitance and magnetocaloric effect \cite{Franco_2012, Parkin_1995, Singh_2015, PhysRevB.89.024405, Gao_2018}. At the level of fundamental physics, a common ingredient seems to be a delicate interplay of spin, charge, and lattice degrees of freedom that leads to various magnetic phase transitions, and the sensitivity of these transitions to external fields or pressure manifests in the form of a technologically useful effect \cite{PhysRevLett.108.127201, Ali_2019, PhysRevB.96.214407}.

Oxides of transition metals are perhaps the most well known materials that are equally interesting from fundamental physics as well as application points of view  \cite{Rao_1989, Kobayashi1998, ALI2020}. Double perovskites (DP) with formula $R_2$BB'O$_6$, where $R$ is rare earth and B, B' are the transition metal (TM) ions, belong to this interesting class of materials \cite{Das_2008, Ragado_2005, Balli_2014, Wei_2014}. The presence of two TM ions leads to multiplet of possibilities in terms of tuning the properties in a desired manner \cite{VASALA20151, Saha-Dasgupta2013, PhysRevB.82.174440}. The coupling between Rare earth and TM network further enriches the low temperature magnetic properties \cite{PhysRevLett.108.127201}. 

MCE remains a topic of immense interest as it opens pathway towards realization of clean and energy efficient magnetic refrigeration technology with clear advantage over the conventional vapor-compression techniques that use potentially harmful chlorofluorocarbon (CFC) gases \cite{PhysRevB.64.144406}.
Recent studies show a giant and reversible MCE in various magnetic materials such as MnFeP$_{0.45}$As$_{0.55}$ \cite{Tegus2002}(magnetic entropy change -$\Delta$S$_M$ = 18.0~J/kg-K for $\Delta$H = 0-5~T), Gd$_5$(Si$_x$Ge$_{1-x}$) \cite{PhysRevLett.78.4494}, Ni-Mn-In \cite{Manosa2010} (adiabatic temperature change $\Delta$T$_{ad}$ = 6.2~K), Gd$_2$NiMnO$_6$ \cite{Krishna_Murthy_2015} (-$\Delta$S$_M$ = 35.5~J/kg-K for $\Delta$H = 0-7~T). The materials which show MCE at low temperatures can be advantageous for cryogenic magnetic cooling to obtain a sub-kelvin temperature as an alternative option of He3/He4 liquid whose prices are constantly increasing. Observation of a significant MCE in DPs opens up possibilities for using the high degree of flexibility available in these oxides to enhance the effect further. We have recently studied the magnetic and magneto-caloric properties of Nd$_2$NiMnO$_6$ and proposed that an interplay of the two magnetic sub-lattices can be used as a control knob to tune the MCE properties of magnetic materials with multiple magnetic sublattices and specifically the DPs~\cite{Ali_2019}.  It would be instructive to test this proposal by varying the moment size on one of the magnetic sublattices to see how the magnetic and MCE properties evolve.  This motivates our present study.

In this work, we present detailed experimental investigations of magnetization and magnetocaloric behavior from 2~K to 300~K on $R_2$NiMnO$_6$ ($R =$ Sm, Gd, Pr, Nd, Tb, and Dy). All studied DPs go through a paramagnetic to ferromagnetic phase transition due to the ordering of Ni$^{2+}$ and Mn$^{4+}$ magnetic ions. The choice of $R$, in addition to affecting the value of T$_{\rm C}$, also influences the magnetization and consequently the MCE behavior as inferred from the nature of low-temperature anomalies for different $R$. We propose a simple explanation for this behavior in terms of a Heisenberg model that takes into account the coupling of the spin on $R$ with the spin on Ni-Mn network, as well as the local spin-orbit coupling on $R$ sites. We support our experimental findings via Monte Carlo simulations and mean-field analysis on the two-sublattice Heisenberg model. The remainder of the paper is organized as follows. In section \ref{experiment}, we present the magnetization measurements for all studied DPs. The low temperature anomalies in magnetization as well as in magnetic entropy change are presented and discussed here. In section \ref{theory}, we present a two-sublattice Heisenberg model for magnetism in DPs and describe the mean-field approximation and Monte-Carlo simulation method to study the model. Further, we discuss the results obtained by these theoretical methods in this section. We summarize and  conclude in section \ref{Conclusion}.

\section{Experimental Results}
\label{experiment}
\begin{figure}
	\centering
	\includegraphics[width=1.0\linewidth, height=0.5\textheight, trim = 0cm 0cm 0cm 0cm]{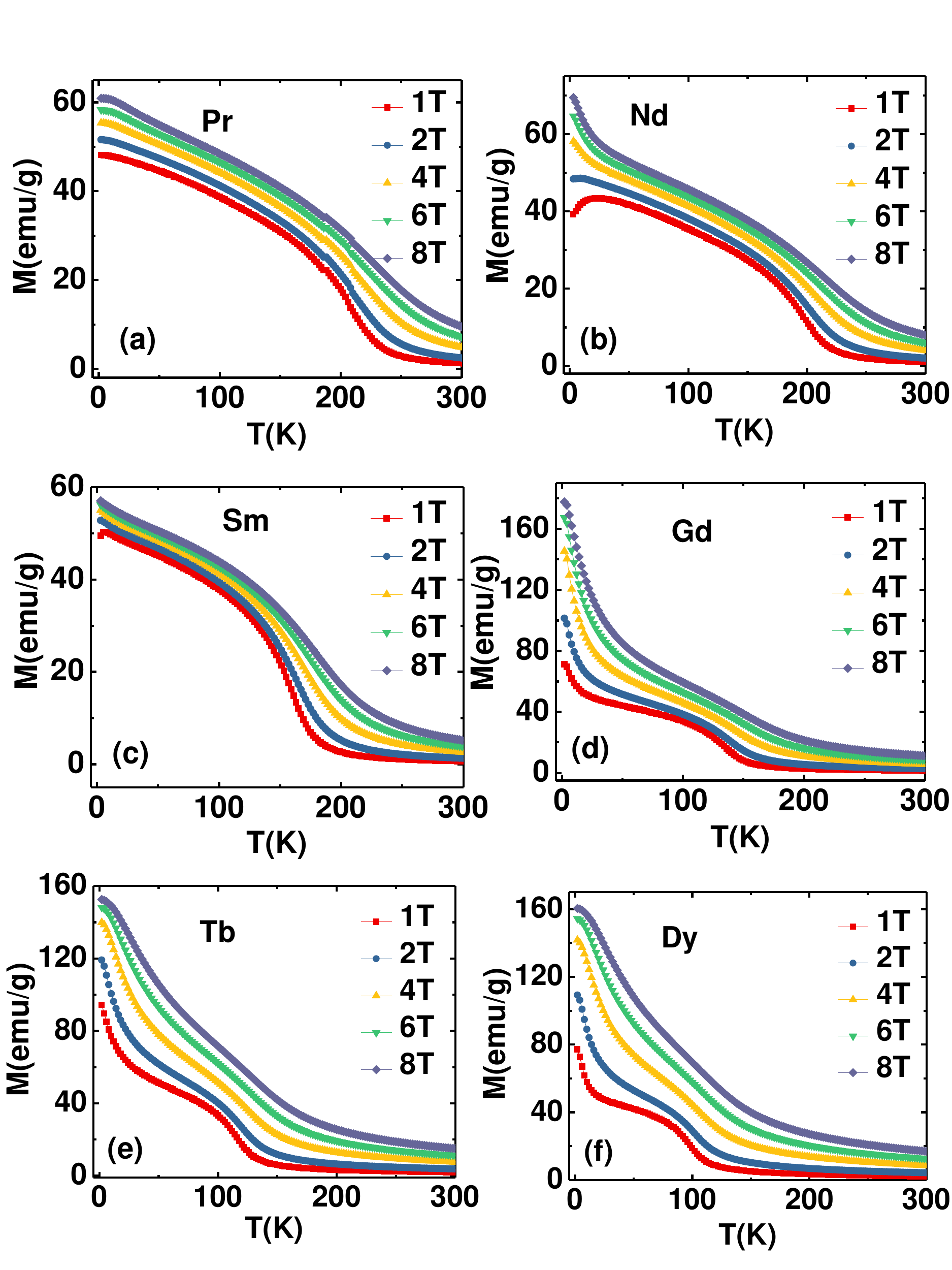}
	\caption{Temperature dependence of the magnetization measured in field-cooled protocol in applied magnetic fields up to $8$~T for all the synthesized $R_{2}$NiMnO$_{6}$ compounds.}
	\label{MT_11}
\end{figure}

Polycrystalline samples of $R_2$NiMnO$_6$ where $R =$ Pr, Nd, Sm, Gd, Tb, and Dy were synthesized using a solid state reaction method as described previously in Ref.~\cite{doi:10.1063/1.5032637, BOOTH20091559}. The phase purity of all DPs was examined by powder X-ray diffraction using a Rigaku Ultima IV Diffractometer. These X-ray diffraction results confirmed that all materials were single phase and crystallized in the expected crystal structure.  Figure \ref{MT_11} shows the temperature dependent magnetization $M(T)$ for all synthesized DPs at different fields as indicated in the plots.

All the DPs show a rapid increase in magnetization $(M)$ at temperatures T$_{\rm C}$ which is associated with the ferromagnetic ordering of Ni$^{2+}$-Mn$^{4+}$ sub-lattice. For higher field this upturn in magnetization moves to higher temperature which is a common feature of a ferromagnetic transition. At lower temperatures, a second anomaly is observed for all the DPs at T$_{\rm 2}$. The values of T$_{\rm C}$ and T$_{\rm 2}$ for all DPs which were studied, are given in Table \ref{T_TC} , which are consistent with the previous reports \cite{CHAKRABORTY201759, BOOTH20091559, PhysRevB.68.064415}.  The nature of the low temperature anomaly depends on the $R$ ion. For $R =$ Pr, Nd, and Sm, there is a downturn in $M$ at T$_{\rm 2}$ when measured in low magnetic fields.  This downturn in $M$ can be suppressed on the application of a magnetic field and changed to an upturn at larger magnetic fields as previously reported \cite{Ali_2019}.  This downturn in magnetization however is not a signature of long range ordering of $R$ ions as demonstrated for Nd$_2$NiMnO$_6$ for which a low temperature heat capacity study did not show any anomaly  \cite{Ali_2019} and microscopic probes like XMCD did not show any ordered moment on Nd$^{3+}$ ions in Nd$_2$NiMnO$_6$ \cite{PhysRevB.100.045122}.  On the other hand, for $R =$ Gd, Tb, and Dy there is an increase in $M$ at T$_{\rm 2}$ at low fields which is enhanced in larger fields.  Taken at face value the above observations seem to indicate that there is antiferromagnetic coupling between the $R$ and Ni-Mn sublattice for $R =$ Pr, Nd, and Sm, while this coupling is ferromagnetic for $R =$ Gd, Tb, and Dy. \cite{doi:10.1063/1.4752262, doi:10.1063/1.4906989, Das_2019}.   However, recent DFT calculations of the exchange constants in Nd$_2$NiMnO$_6$ have shown that the magnetic exchange between the Nd spin and the Ni-Mn sublattice is ferromagnetic \cite{PhysRevB.100.045122}.  The downturn in the magnetization can then be understood by realizing that for less than half-filled $4f$ shells, the orbital moment of the $R$ is oppositely aligned to it's spin moment and is larger in magnitude than the spin moment.  Therefore, although the spin is coupled ferromagnetically to the Ni-Mn sublattice, the orbital moment is opposite to the Ni-Mn sublattice resulting in a downturn in the magnetization for $R =$ Pr, Nd, and Sm.  At large enough magnetic fields, it becomes energetically favorable for the net effective magnetic moment to align with the field, which leads to a switching of the magnetization anomaly at T$_{\rm 2}$ from a downturn to an upturn as field is increased.  For $R = $ Gd, Tb, and Dy the orbital moment is in the same direction as the spin moment and so the ferromagnetic coupling between the spin of $R$ and the Ni-Mn sublattice leads to an upturn at T$_{\rm 2}$.  
 
\begin{figure}
	\centering
	\includegraphics[width=1.0\linewidth, height=0.5\textheight, trim = 0cm 0cm 0cm 0cm]{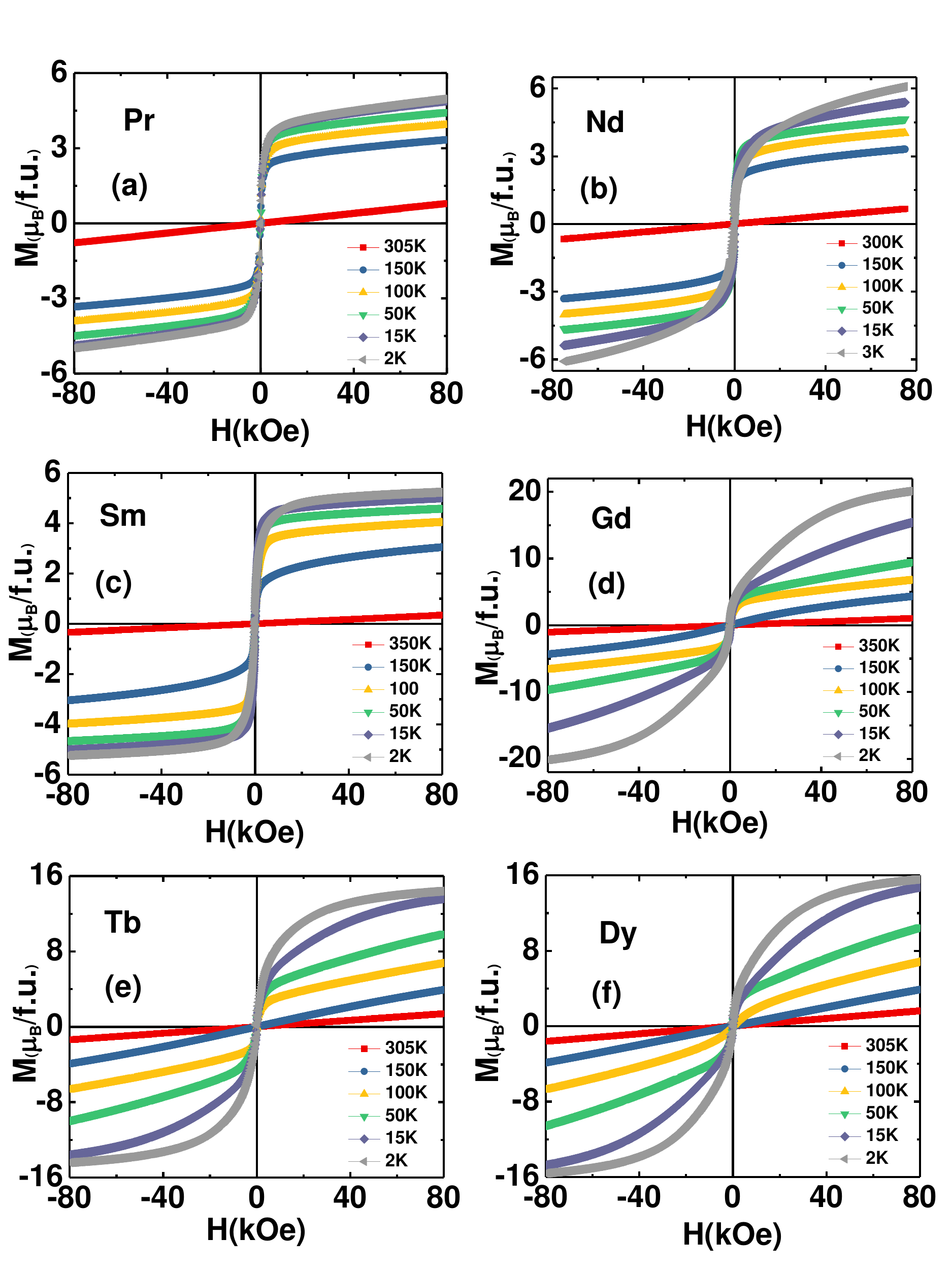}
	\caption{Isothermal curves of magnetization verses magnetic field for all the $R_{2}$NiMnO$_{6}$ at various temperatures indicated in the plots.}
	\label{MH_1}
\end{figure}

\begin{table*}
\begin{center}
\caption{Theoretically estimated effective magnetic moment of rare earth ions (m) \cite{ref999921680802121}, ferromagnetic transition temperature (T$_{\rm C}$), anomaly at lower temperature (T$_{\rm 2}$), and theoretically estimated and experimentally observed saturation magnetization (M$_S$) for all DPs $R_2$NiMnO$_6$.}
\vspace{0.5cm}
\label{T_TC}
\setlength\extrarowheight{12pt}
\setlength{\tabcolsep}{16pt}
\begin{tabular}{c c c c c c}
\hline
\hline
{$R_2$NiMnO$_6$} & {m($\mu_B$)} & {T$_{\rm C}$(K)} & {T$_2$(K)} & {M$_S$(theoretical)} & {M'$_S$(observed)} \\ \hline
Dy$_2$NiMnO$_6$ & 10.63 &  95 & 21 & 25.00 & 15.90 \\ \hline
Tb$_2$NiMnO$_6$ &  9.72 & 113 & 15 & 23.00 & 15.45 \\ \hline
Gd$_2$NiMnO$_6$ &  7.94 & 130 & 20 & 19.00 & 18.90 \\ \hline
Nd$_2$NiMnO$_6$ &  3.62 & 195 & 50 & 11.54 &  6.02 \\ \hline
Pr$_2$NiMnO$_6$ &  3.58 & 213 & 27 & 11.40 &  5.25 \\ \hline
Sm$_2$NiMnO$_6$ &  0.85 & 158 & 20 &  6.43 &  5.10 \\ \hline
\hline 
\end{tabular}
\end{center}
\end{table*}

\begin{figure}[h]
	\centering
	\includegraphics[width=1.0\linewidth, height=0.5\textheight, trim = 0cm 0cm 0cm 0cm]{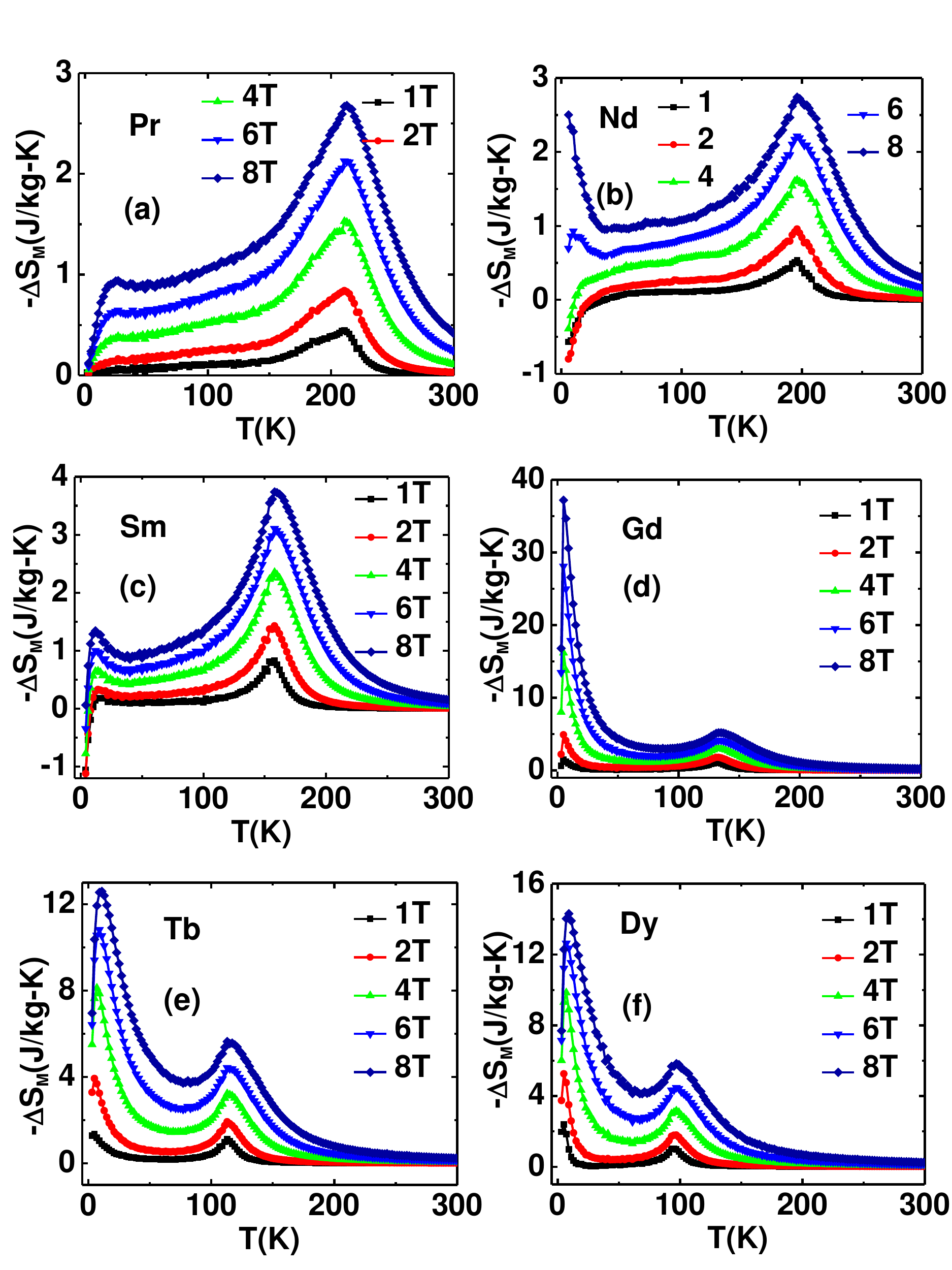}
	\caption{Temperature dependence of magnetic entropy change (-$\Delta$ S$_M$(T)) from 2K to 300K at applied magnetic fields up to 8T for all the $R_{2}$NiMnO$_{6}$ compounds.}
	\label{MCE_1}
\end{figure}

Figure \ref{MH_1} shows the isothermal magnetization $M(H)$ of all $R_{2}$NiMnO$_{6}$ at different temperatures as indicated in plots. All of the DPs were found to show a negligible hysteresis in magnetization down to the lowest measured temperature $2$~K, indicating soft ferromagnetic behavior which is a desirable feature for the magnetic refrigeration techniques. At 300~K, in the paramagnetic state, $M(H)$ for all the DPs varies linearly. Below T$_{\rm C}$ but above T$_{\rm 2}$, $M(H)$ shows tendency to saturate but with a weak linear increase at large $H$ as expected in the ferromagnetic state of Ni-Mn, while the $R$ ions are still paramagnetic.  Below T$_{\rm 2}$ the $M(H)$ curves are complicated by the different behaviour for DPs with less than ($R =$ Pr, Nd, Sm) and more than or equal to ($R =$ Gd, Tb, Dy) half filled $4f$-shells. For $R =$ Pr, Nd, and Sm, the $M(H)$ at low fields is lower than for $T > T_2$ as seen in the downturn in the $M(T)$ below T$_{\rm 2}$ due to the orbital momnent being anti-aligned with the Ni-Mn sublattice. At higher fields for which the downturn in $M(T)$ changes to an upturn due to field induced aligning of the orbital moment with Nio-Mn sublattice, the $M(H)$ increases and shows a tendency to saturate.  However, a weak linear increase in $M(H)$ with field remains down to $T =- 2$~K, suggesting that the full magnetic moment of $R$ has still not aligned with the field.  The values of observed saturation magnetization (M'$_S$) at $T = 2$~K, $H = 9$~T, and the theoretically expected saturation magnetization (M$_S$) if all ($R$, Ni, and Mn) magnetic moments have saturated are also given in Table \ref{T_TC}. The value of observed saturated magnetization for all $R_{2}$NiMnO$_{6}$ is lower than the theoretically predicted saturation magnetization except for Gd$_{2}$NiMnO$_{6}$ probably because Gd has no orbital moment (L). 
\begin{table*}
	\begin{center}
		\caption{Theoretically estimated effective magnetic moment of rare earth ions (m), -$\Delta$S$_M$ at T$_{\rm C}$, $\delta$T$_{FWHM}$, RCP, -$\Delta$S'$_M$ below T$_S$ for all DPs $R_2$NiMnO$_6$ at an applied field of 8~T.}
		\vspace{0.5cm}
		\label{T_M1}
		\setlength\extrarowheight{12pt}
		\setlength{\tabcolsep}{16pt}
		\begin{tabular}{c c c c c c}
			\hline
			\hline
			{$R_2$NiMnO$_6$} & {m($\mu_B$)} & {-$\Delta$S$_M$(J/kg-K)} & {$\delta$T$_{FWHM}$} & {RCP(J/kg)} & {-$\Delta$S'$_M$(J/kg-K)} \\ \hline
			Dy$_2$NiMnO$_6$ & 10.63 &  5.82 & 67  & 389.47 & 14.31 \\ \hline
			Tb$_2$NiMnO$_6$ &  9.72 &  5.63 & 71  & 400.41 & 12.59 \\ \hline
			Gd$_2$NiMnO$_6$ &  7.94 &  5.16 & 72  & 371.00 & 37.18 \\ \hline
			Nd$_2$NiMnO$_6$ &  3.62 &  2.95 & 100  & 295.00 &  2.50 \\ \hline
			Pr$_2$NiMnO$_6$ &  3.58 &  2.67 & 109 & 291.43 &  0.94 \\ \hline
			Sm$_2$NiMnO$_6$ &  0.85 &  3.73 & 74 & 276.62 &  1.34 \\ \hline
			\hline 
		\end{tabular}
	\end{center}
\end{table*}

These field and temperature dependent magnetic responses also manifest themselves in the magnetic entropy change. The isotherms of magnetization $M(H)$ were collected over a broad temperature range from 2~K to 300~K at temperature intervals of 2~K.  Using these $M(H)$ data, the change in magnetic entropy ($\Delta$S$_M$) can be calculated from Maxwell's thermodynamic relation \cite{doi:10.1146/annurev-matsci-062910-100356}:

\begin{equation}
\Delta S_M (H,T)=\int_{0}^{H} \left( \dfrac{dM}{dT}\right) _{H'} dH' 
\label{Maxwell_eqn}
\end{equation}

Figure \ref{MCE_1} show the magnetic entropy change (-$\Delta$S$_M$) versus temperature at different applied magnetic fields up to $8$~T for all the $R_{2}$NiMnO$_{6}$ compounds. All the compounds show two prominent features. The first one at the ferromagnetic phase transition at T$_{\rm C}$ and second one at lower temperature. The value of -$\Delta$S$_M$ is positive for all the DPs across T$_{\rm C}$ and it's magnitude increases with field, which is consistent with ferromagnetic ordering of the Ni-Mn sublattice at T$_{\rm C}$ as indicated in the magnetization versus $T$ and $H$.  When the temperature is lowered, the behavior of -$\Delta$S$_M$ like $M(T)$, is of two kinds.  For $R =$ Pr, Nd, and Sm, the -$\Delta$S$_M$ monotonically decreases below T$_{\rm C}$ and even becomes negative for $R$ = Nd and Sm at low fields. On increasing the applied magnetic field, the negative -$\Delta$S$_M$ is suppressed and a positive peak or anomaly appears at low temperatures and high fields.  This can be understood interms of the anti-alignment of the total effective moment of the $R$ ions with the Ni-Mn sublattice below T$_{\rm 2}$ at low fields.  So the magnetic entropy is high.  At higher fields, the $R$ moment switches and aligns with the Ni-Mn sublattice and also the applied magnetic field, resulting in a low magnetic entropy like in the case of the ferromagnetic transition at T$_{\rm C}$.  Gd$_{2}$NiMnO$_{6}$, Tb$_{2}$NiMnO$_{6}$, and Dy$_{2}$NiMnO$_{6}$ show a positive anomaly in -$\Delta$S$_M$ below T$_{\rm 2}$ since the effective moment of $R$ is already tending to align with the Ni-Mn sublattice.  This alignment simply increases as higher magnetic fields are applied resulting in an enhancement in the magnitude of -$\Delta$S$_M$.  For $R =$ Gd, Tb, and Dy the value of -$\Delta$S$_M$ is very large at low temperatures, particularly for $R =$ Gd, which make them potentially useful for low temperature magnetic refrigeration. 

Another quantity which is used to quantify the usefulness of MCE materials is called relative cooling power (RCP) which is defined as:

\begin{equation}
{\rm RCP} =  -\Delta S_M(T,H) \times \delta T_{FWHM}
\label{RCP}
\end{equation}     
where $\delta$T$_{FWHM}$ is the full width at half maxima of -$\Delta$S$_M$ for a specific value of applied magnetic field.  Values of -$\Delta$S$_M$ (at T$_{\rm C}$), $\delta$T$_{FWHM}$, RCP, and -$\Delta$S'$_M$ (at low temperature below T$_2$) were calculated for all DPs at $8$~T and are given in Table \ref{T_M1}. 

\begin{figure}
	\centering
	\includegraphics[width=1.0\linewidth, height=0.3\textheight, trim = 0cm 0cm 0cm 0cm]{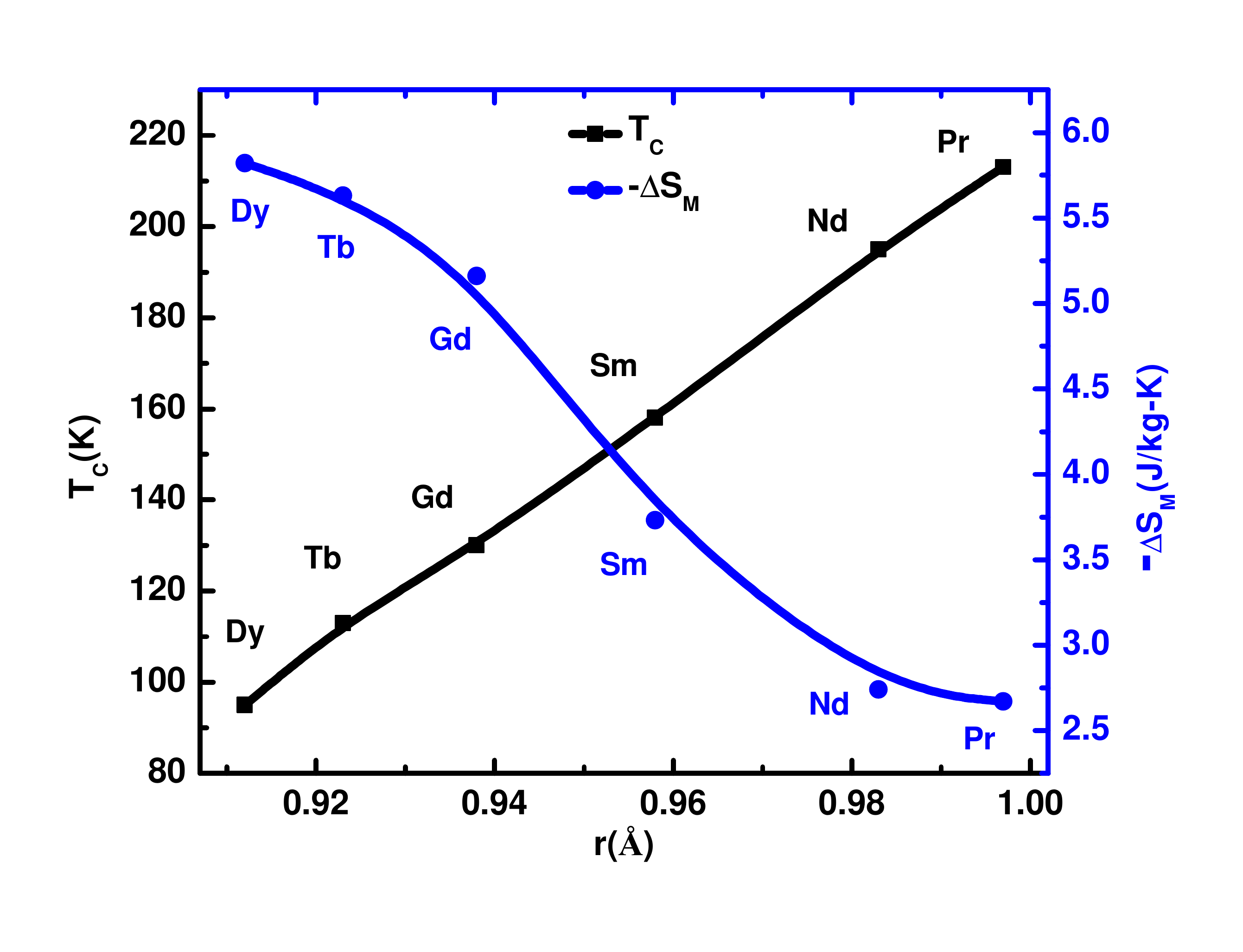}
	\caption{Ferromagnetic transition temperature (T$_{\rm C}$) and magnetic entropy change (-$\Delta$ S$_M$(T)) as a function of the ionic radii of the rare earth ion for all the $R_{2}$NiMnO$_{6}$ compounds.}
	\label{Comp_1}
\end{figure}

Figure \ref{Comp_1} shows the evolution of $T_{\rm C}$ and -$\Delta$S$_M$ at $T_{\rm C}$ with the ionic radius of $R^{3+}$ in $R_{2}$NiMnO$_{6}$.  We find that the ferromagnetic phase transition temperature $T_{\rm C}$ increases linearly and -$\Delta$ S$_M$ at $T_{\rm C}$ decreases monotonically with the ionic size of $R^{3+}$ for all $R_{2}$NiMnO$_{6}$.  The decrease in $T_{\rm C}$ in going from Pr to Dy is understood to arise from a decrease in the Ni-O-Mn bond angle which leads to a decrease of the super-exchange strength \cite{doi:10.1143/JPSJ.67.4218}.  Additionally from Table \ref{T_M1} we observe that the paramagnetic background of rare earth ions spreads the magnetic entropy over a large temperature range across $T_{\rm C}$.         

\section{Theoretical Results}
\label{theory}
\subsection{Heisenberg Model for $R_{2}$NiMnO$_{6}$}
\noindent
The experimental data discussed above presents intriguing magnetic behavior for Ni-Mn DPs. What is particularly interesting is the dramatically different magnetic response for different $R$ at low temperatures. In order to comprehend the experimental observations, we propose a simple Heisenberg model on a body-centered cubic (BCC) lattice that takes into account spin degrees of freedom on B sites and both spin and orbital degrees of freedom on $R$ site. The model is specified by the Hamiltonian, 
\begin{eqnarray}
{\cal H} &=& -J_{1} \sum_{ \langle ij \rangle} {\bf S}_i^{\textrm {Mn}} \cdot{\bf S}_j^{\textrm {Ni}}
- J_{2} \sum_{\langle ij \rangle} {\bf S}_i^{\textrm {R}} \cdot{\bf S}_j^{\textrm {Mn}} \nonumber \\
&&- J'_{2} \sum_{\langle ij \rangle}{\bf S}_i^{\textrm {R}} \cdot{\bf S}_j^{\textrm {Ni}} + \lambda \sum_{i\in \textrm A} {\bf L}^i_{\textrm {R}} \cdot {\bf S}^{\textrm {R}}_{i}   \nonumber \\
&&- g_{s} H \sum_{ i\in \textrm A, \textrm B} S_{i,\textrm z} - g_{L} H \sum_{ i\in \textrm A}  L_{i, \textrm z} . \label{heisen-perov}\end{eqnarray} 
\noindent
In the above, ${\bf S}_i$ with the appropriate superscript denotes the spin of the relevant Ni, Mn or $R$ ions, and ${\bf L}_i^{R}$ denotes the orbital angular momentum on the rare earth ion.
The model explicitly considers two coupled sublattices A and B. Spins on $R^{3+}$ ions on sublattice A are coupled to those on sublattice B of Mn$^{4+}$ and Ni$^{2+}$ \cite{}. In the model, the first three terms have the summation over nearest neighbors (nn) as indicated by angular brackets. $i\in$ A (B) represents lattice sites on A (B) sublattices. $J_{1}$ is the ferromagnetic coupling between Mn$^{4+}$ and Ni$^{2+}$ and $J_{2}(J'_{2})$ is the ferromagnetic coupling of R$^{3+}$ to Mn$^{4+}$ (Ni$^{2+}$). The strength of spin-orbit coupling in rare earth ions is represented by $\lambda$. $H$ symbolizes the strength of uniform external field which couples to the $z$ components of the spin and orbital magnetic moments. We use $g_{s}= 2$ and $g_{L} = 1$ as the  spin and orbital g-factors. In the BCC lattice, Mn$^{4+}$$(S=\frac{3}{2})$ couples to 6 Ni$^{2+}$ $(S=1)$ and vice versa.  Ni$^{2+}$ couples to 8 $R^{3+}$ and similarly  coordination number of Mn$^{4+}$ with $R^{3+}$ is 8. $J_{1}=1$ sets the energy scale for the model. For simplicity we assume  $J'_{2} = J_{2}$, and different R ions are parameterized by different choices of $J_{2}$ and $\lambda$, with the choice $J_{1} >> J_{2}$, $\lambda$ inspired by the experimental data.

\subsection{Methodology}
\noindent 
{\bf Mean-Field Approximation}:The approximation proceeds by rewriting the model as a sum of three effective single-spin Hamiltonians as,
\begin{eqnarray}
{\cal H}_{MF}& =& -g_{s} \sum_{\textrm {Mn}} {\bf {S}}^{\textrm {Mn}}\cdot {\textit {\textbf{H}}}^{\textrm {Mn}}_{eff}-g_{s} \sum_{\textrm {Ni}} {\bf S}^{\textrm {Ni}}\cdot {\textit {\textbf{H}}}^{\textrm {Ni}}_{eff} \nonumber \\
&&-g_{s} \sum_{R} {\bf S}^{\textit R}\cdot {\textit {\textbf{H}}}^{\textit R}_{eff} -g_{L} \sum_{\textit R} {\bf L^{\textit R}}\cdot {\textit {\textbf{H}}}^{L}_{eff}.
\label{mean_heff}
\end{eqnarray}
The effective magnetic field or the molecular field for the spins on different inequivalent sites is given by,
\begin{eqnarray}
{\textit {\textbf{H}}}^{\textrm {Mn} (\textrm{Ni})}_{eff} &=& {\textit {\textbf{H}}} + 6 \frac{J_{1} \langle {\bf S}^{\textrm {Ni}(\textrm {Mn})} \rangle}{g_{s}} + 8 \frac{J_{2} \langle {\bf S}^{\textit {R}} \rangle}{g_{s}} %
\nonumber \\
&& = {\textit {\textbf{H}}} + 6 \frac{J_{1} {\textit{\bf M}^{\textrm{Ni}(\textrm{Mn})}}}{g^2_{s}} + 8 \frac{J_{2} {\textit{\textbf M}_{s}^{\textit{R}}}}{g^2_{s}}.
\label{heff}
\end{eqnarray}
In the above, we replace each spin operator ${\textbf{S}}^{\textrm{Ni}(\textrm{Mn})}$ by its average value $\langle {\textbf S}_{i}^{\textrm{Ni}(\textrm{Mn})} \rangle$ in the mean-field spirit, and the sublattice-resolved magnetization is then given by, ${\textit{\textbf M}}^{\textrm{Ni}(\textrm{Mn})} = g_{s}\langle {\textbf S}^{\textrm{Ni}(\textrm{Mn})} \rangle$. 
The subscript in ${\textbf {M}}_s$ is necessary for the $R$ sites in order to differentiate between spin and orbital contributions. Similarly we find,
\begin{eqnarray}
{\textit {\textbf{H}}}^{R}_{eff} &=&  {\textit {\textbf{H}}} + 4 \frac{J_{2}}{g^2_{s}}[{\textit{\textbf M}}^{\textrm {Ni}} + {\textit{\textbf M}}^{\textrm{Mn}}] - \lambda \frac{{\textit{\textbf M}}_{L}^{R}}{g_s g_{L}}
\end{eqnarray}
In the above equation, we used $M^{R}_{L} = g_{L}\langle L_{R} \rangle$. Also,
\begin{eqnarray}
{\textit {\textbf{H}}}^{L}_{eff} &=&  {\textit {\textbf{H}}} - \lambda \frac{{\textit{\textbf M}}_{s}^{R}}{g_{s} g_{L}}
\end{eqnarray}

Following the textbook procedure of writing statistical average $\langle S \rangle$ in terms of the partition function, we get magnetization on different sublattices as,
\begin{eqnarray}
M^{i}_{s} = g_{s} S^{i} B_{S^{i}}(x^{}),~~~~~~~~~~ \forall ~i\in \{ {\textrm{Mn}}, {\textrm{Ni}}, R \} \nonumber
\label{mf_mag}
\end{eqnarray}
\begin{eqnarray}
M^{R}_{L} = g_{L} L^{R} B_{L^{R}}(y^{}),~~~~~~~~~~
\end{eqnarray}
with $x^{}=\frac{g_{s}S^{i}H_{eff}^{S^{i}}}{T}$ and $y^{}=\frac{g_{L}L^{R}H_{eff}^{L^{R}}}{T}$, where 
$S^{i}$ and  $L^{R}$ take the maximum value of the respective spin and orbital angular momenta and the Brillouin function is given by, 
\begin{eqnarray}
B_{p}(x^{}) = \frac{2p+1}{2p} \coth( \frac{2p+1}{2p} x^{})- \frac{1}{2p}\coth(\frac{x^{}}{2p}).
\end{eqnarray}
This leads to a system of coupled equations which are then solved self-consistently.  The total magnetization is calculated as, $M = M^{{\textrm{Mn}}} + M^{{\textrm{Ni}}} + M^{R}_s + M^{R}_{L}$. We also determine 
the change in entropy $(\Delta S_{M})$ as a function of temperature via,

\begin{equation}\label{entropy}
\Delta S_{M}(T_{i},H) =  \small \sum \limits_{j=1}^{p} \frac{M(T_{i+1},H_{j})- M(T_{i-1},H_{j})}{T_{i+1}-T_{i-1}} (H_{j+1}-H_{j}).
\end{equation} 
\normalsize
\noindent
$\Delta S_{M}(T_{i},H)$ is thus the change in magnetic entropy at a certain discrete value $T_{i}$ of temperature. $H_{j}$ arises from uniform discretization of interval $ [0,H]$ such that, $H_1=0$ and $H_p=H $. The above equation is the discrete version of continum equation, Eq. (\ref{Maxwell_eqn}).\\ \par

\noindent
{\bf Classical Monte Carlo Method}:
Since the mean-field approximations discussed above completely ignores statial correlations, we use Monte-Carlo simulations to ensure that the low-temperature magnetic behavior is correctly captured.
We employ the standard Markov chain Monte Carlo method with the Metropolis algorithm \cite{newman-1999}. Typical simulated lattice size is $2 \times 16^{3}$ sites, with periodic boundary conditions. For thermal equilibration and averaging of quantities, we use $\sim 10^{5}$ Monte Carlo steps each. Alongside, the change in entropy $(\Delta S_{M})$ as a function of temperature  (Eq. \ref{entropy}), the {\textit z} component of total magnetization is calculated as, 
\begin{equation}
M_z = \frac{1}{N}  \left \langle g_s \sum_{i\in A,B} {S}^{i}_{z} \right \rangle +  \frac{1}{N}  \left \langle g_L \sum_{i\in A} {L}^{i}_{z} \right \rangle,
\end{equation}
where $N$ is total number of spins and the angular bracket denotes the thermal average over Monte Carlo generated equilibrium configurations.
\subsection{Results and Discussions}
\begin{figure}[!htbp]
	\includegraphics[width=1.0\columnwidth,angle=0, clip='True']{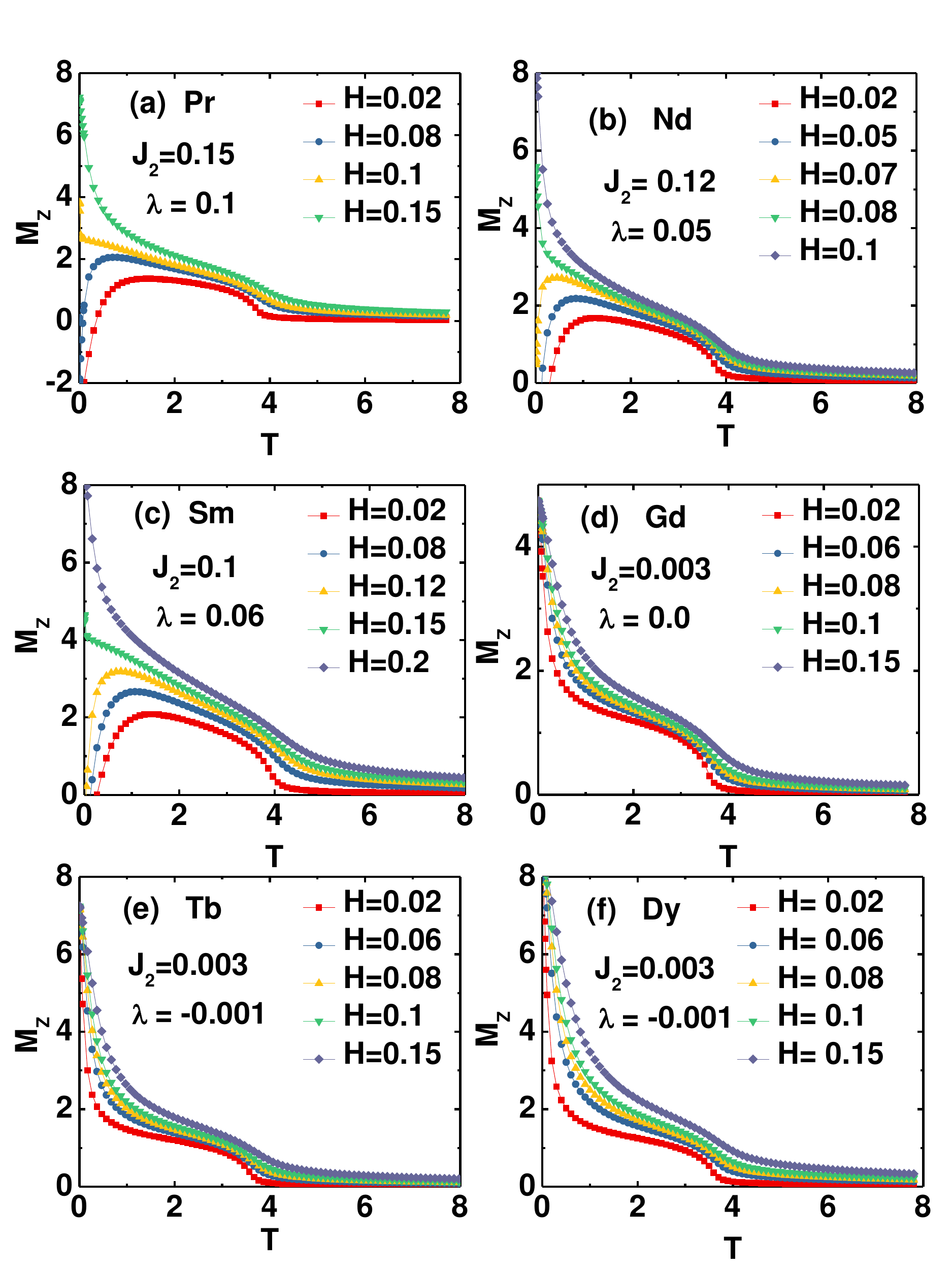}
	\caption{Monte Carlo results for magnetization ($M_{z}$) as function of temperature ($T$) at various strengths of external field for $S^{\textit {R}}$ and $L^{\textit {R}}$ values corresponding to (a) $R=$ Pr, (b) $R=$ Nd, (c) $R=$ Sm, (d) $R =$ Gd, (e) $R=$ Tb, and (f) $R=$ Dy. We set $J_{1} = -1.0$ as the energy scale. $J_{2}$ and $\lambda$ are model parameters.} 
	\label{fig1}
\end{figure}

\noindent
We begin with the results obtained from the classical Monte Carlo simulations using Hamiltonian Eq. (\ref{heisen-perov}). First, we discuss the variation of magnetization ($M_{z}$) with temperature at various values of magnetic field strength ($H$). In Fig. \ref{fig1} we find that for all choices of $R$, there is a smooth rise in $M_{z}$ as the temperature is lowered. This indicates the presence of a second-order phase transition from paramagnetic to ferromagnetic state below a critical temperature (T$_C \approx 4$). The transition is due to ferromagnetic coupling between the magnetic moments of Ni-Mn network in the paramagnetic background of rare earth moments \cite{Pal_2019}. The interesting behavior is noted at low temperature where $M_{z}$ shows an upturn or a downturn depending on the choice of $R$. The experimental data clearly indicates an upturn for $R =$ Gd, Tb, Dy, and a weak downturn for $R =$ Sm, Pr, Nd. Since Gd does not support any orbital contribution to magnetism, the upturn in magnetization for Gd confirms that the coupling $J_2$ is ferromagnetic. The sign of the coupling between spin and orbital moments on R ions, however, depends on the filling of f-shells. Indeed, in accordance with the Hund's rule, the orbital and spin moments will antialign (align) for less (more) than half filled band. If this argument borrowed from atomic physics holds then we expect that a simple change in sign of $\lambda$ should capture the experimental behavior of magnetization across the entire rare earth series.
Indeed, using the magnitude of ferromagnetic $J_2$ and $\lambda$ as model parameters, our Monte Carlo results describe the experimental data very well. For $R =$ Sm, Pr, Nd,  at lower values of $H$, we notice a downturn in magnetization much below $T_{C}$ (see Fig. \ref{fig1} (a), (c), (d)).  The downturn at low temperatures indicates an antiferromagnetic correlation of rare earth moments with Ni-Mn moments. However, the origin of this distinct behavior lies in the spin-orbit coupling. Once the applied field becomes stronger than the antiferromagnetic tendency, magnetization begins to show an upturn. The simulations for $R =$ Gd, Tb, Dy are also consistent with the experiments, and display an upturn (see Fig. \ref{fig1} (b), (e), (f)). \\

\begin{figure}[!htbp]
	\centering
	\includegraphics[width=1.0\columnwidth,angle=0, clip='True']{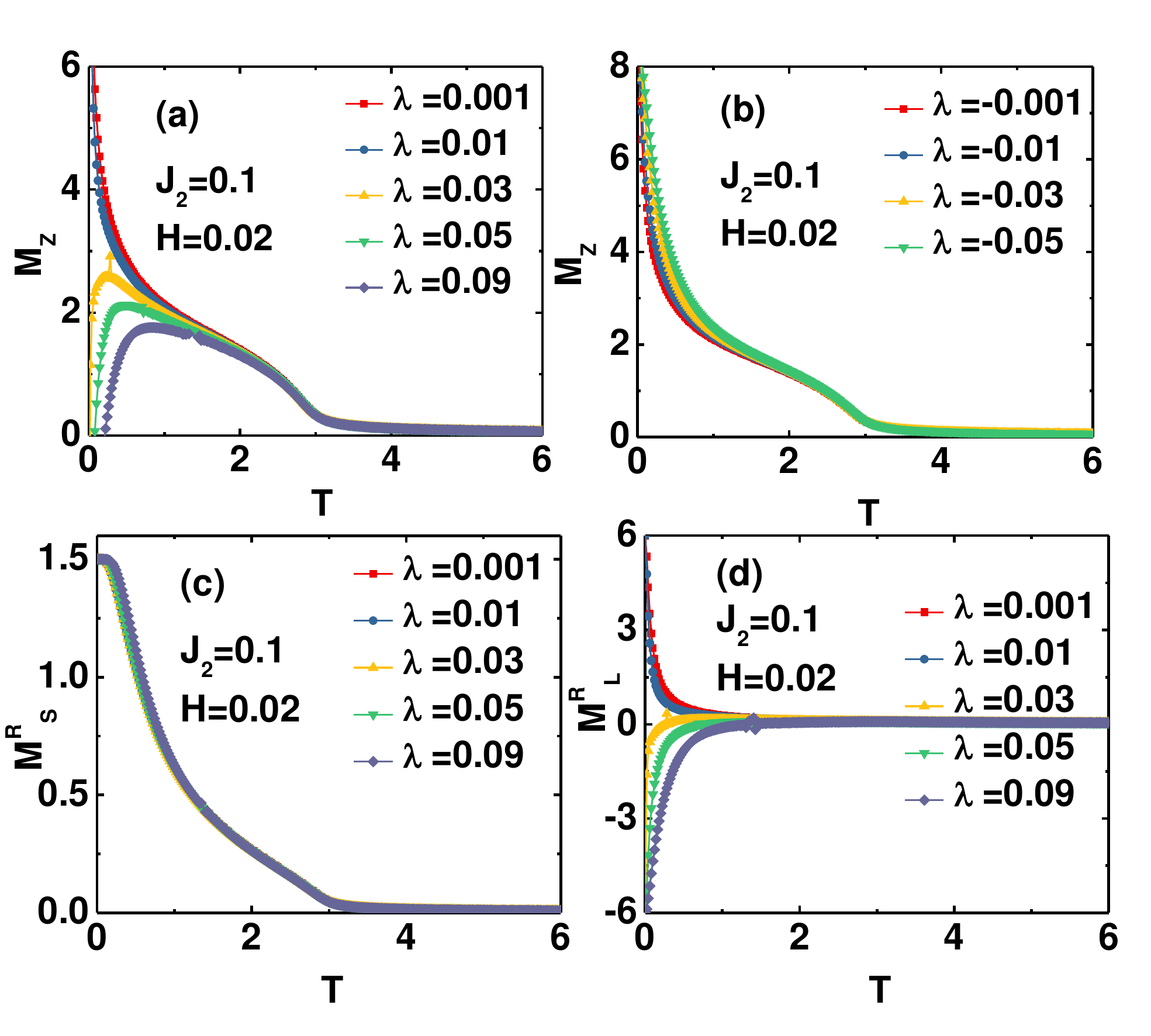}
	\caption{Mean-field results for $L^R = 6$ corresponding to Nd at $H = 0.02$. (a)-(b) The variation of magnetization ($M_{z}$) as function of temperature ($T$) for different strengths and signs of spin-orbit coupling ($\lambda$). (c) and (d) Temperature dependence of spin and orbital magnetization on $R$ sublattice for positive $\lambda$. The results are obtained for $J_1=1$ and $J_2 = 0.1$.} 
	\label{fig4}
\end{figure}

\indent
We have seen how the Monte Carlo data resembles the experimental data for all choices of $R$ by simply tuning the values of $\lambda$ and $J_2$. We clarify further the underlying key idea by even simpler mean-field calculations described in the methods section. 
In order to emphasize the essential features, 
we set $J_{1} = 1$ and $J_{2} = 0.1$ for the mean field calculations, and use $L^R = 6$ corresponding to $R=$Nd.
The low-temperature behavior of magnetization obtained within mean-field calculations (see Fig. \ref{fig4}(a)-(b)) confirms that the Hund's rule controling the antialignment or alignment of spin and orbital moments on the rare-earth ion can explain the qualitatively distinct behavior for different $R$. Also from Fig. \ref{fig4} (c), we find that for all values of $\lambda$, $R$ spin moments begin to allign ferromagnetically with the Ni-Mn sublattice as the temperature is lowered starting at T$_{\rm C}$. On the other hand, the orbital moments on $R$ remain disordered until the temperature scale becomes lower than the energy scale of the spin orbit coupling $\lambda$. For a positive sign of $\lambda$, the orbital and spin moments on $R$ antialign (see Fig. \ref{fig4} (d)). Similar mean-field results are obtained by using the $L_R$ values corresponding to Sm and Pr (not shown). As mentioned earlier, the AFM spin-orbit interaction on R ions can be justified on the basis of Hund's rule. The $R^{+3}$ ions with $R =$ Sm, Pr, Nd have 4f subshells which are less than half-filled. According to Hund's rule, if the subshell is less than half-filled, then the stable configuration has the minimum total angular momentum ($J$) value. The minimum $J$ is possible only if the orbital moment of rare-earth ion aligns oppositely to the spin magnetic moment. \\
\indent
The low temperature downturn in $M_{z}$, for instance in Nd$_{2}$NiMnO$_{6}$ can be understood as follows. The orbital moment of Nd$^{+3}$, $L = 6$, is much larger than the corresponding spin value. As the orbital moment and the spin moment of Nd$^{+3}$ are oppositely aligned, the total moment is in the direction of orbital moment. Additionally, the Nd spin moments on A sublattice is ferromagnetically coupled to Ni-Mn spin moments on B sublattice. Therefore, the total magnetic moment of the Nd sublattice is antialigned to the total magnetic moment of the Ni-Mn sublattice leading to a decrease in magnetization at low temperatures. With increasing magnetic field the Hund's rule energy is compensated by the external field leading to an 
upturn in $M_{z}$. Similar physics persists in Sm$_{2}$NiMnO$_{6}$ and Pr$_{2}$NiMnO$_{6}$ double perovskites. On the other hand, in double perovskites with $R =$ Tb, Dy we find that ferromagnetic coupling between the rare earth orbital moment and the spin moment captures the upturn in $M_{z}$ in the low-temperature regime. The alignment of spin and orbital contributions to magnetization in case of Tb and Dy is again attributed to Hund's rules. As the 4f ions have more than half-filled f orbitals so the minimum energy state must have maximum $J$. The maximum $J$ is obtained when the orbital moment is parallel to the spin magnetic moment of rare-earth ion. The upturn can thus be related to the large total magnetic moment of rare-earth ion on A sublattice coupling ferromagnetically to B sublattice. $R =$ Gd represents a special case where the orbital contribution does not exist, and therefore it also serves as a checkpoint for the choice of sign of $J_2$ used in our model. Our mean-field results on Gd$_{2}$NiMnO$_{6}$ are consistent with a previous mean-field study on this material~ \cite{doi:10.1143/JPSJ.67.4218}. \par
\begin{figure}[!htbp]
\centering
\includegraphics[width=1.0\columnwidth,angle=0, clip = 'True']{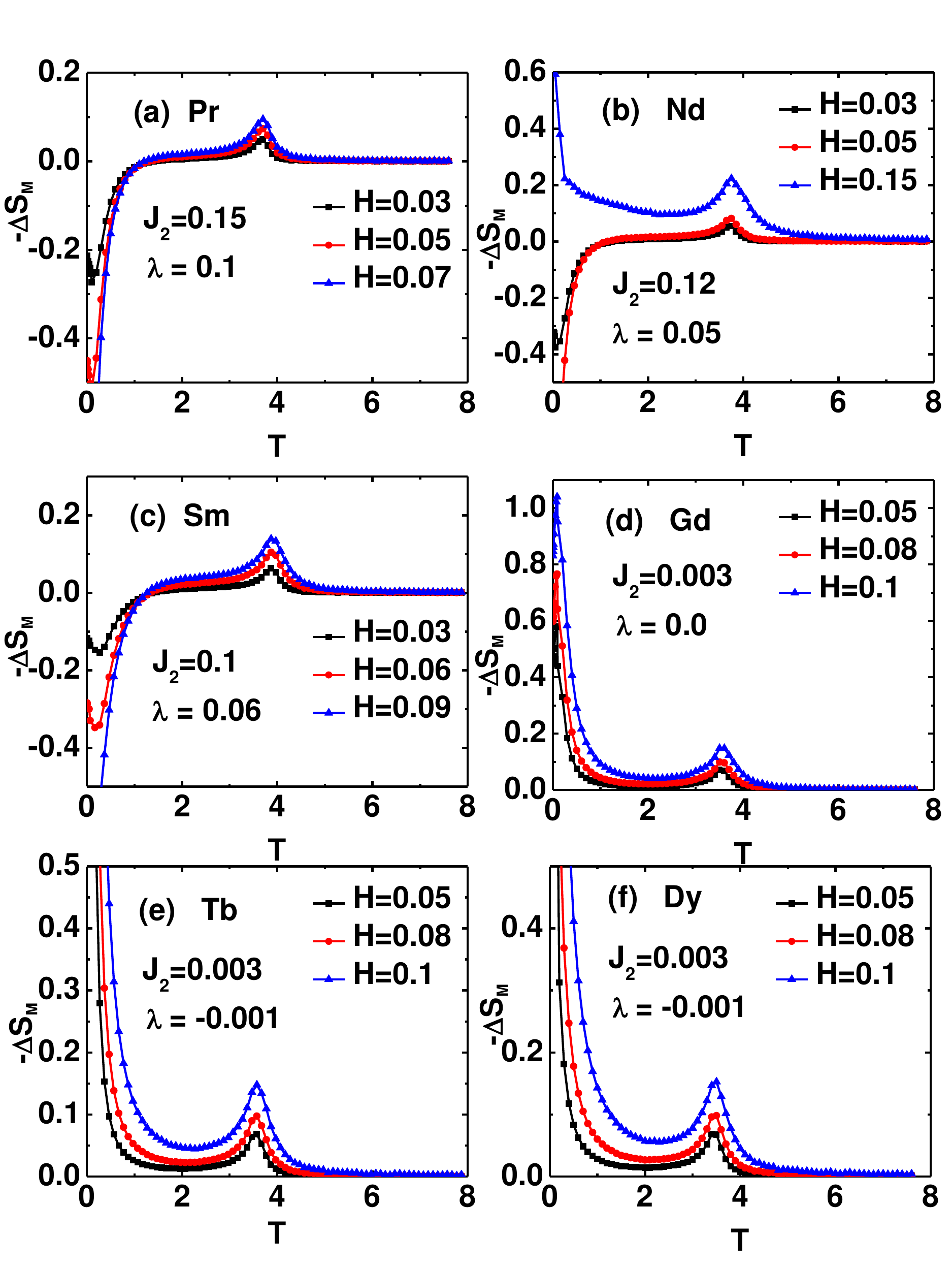}
\caption{Monte Carlo simulation results for change in magnetic entropy ($-\Delta S_{M}$) as function of temperature ($T$) at various strengths of external field for $S_R$ and $L_R$ values corresponding to (a) $R=$ Pr, (b) $R=$ Nd, (c) $R=$ Sm, (d) $R =$ Gd, (e) $R=$ Tb, and (f) $R=$ Dy.} 
	\label{fig3}
\end{figure}

From the above discussion, we conclude that the low temperature behavior in DPs with $R =$ Sm, Nd, Gd, Tb, Dy observed in our experiments is qualitatively captured via a minimal model that explicitly considered the orbital degree of freedom on $R$ sites. The case of $R=$ Pr presents a slight disagreement as we do not find a downturn in $M_{z}$ at low temperatures in our low-field data. One possibility for this could be a smaller spin-orbit coupling in $R=$ Pr that can be easily overcome even by a small magnetic field. \\  
\indent
Further, we present results of change in entropy as a function of temperature obtained via our Monte-Carlo simulations. The results are shown in Fig. \ref{fig3}. Given that $\Delta S_{M}$ is a quantity derived from $M_z(H,T)$, it is not surprizing that except for $R=$ Pr the results match well with the data reported in our experiments.
We obtain $-\Delta S_{M}>0$ near the ferromagnetic transition of Ni-Mn sublattice for all choices of $R$. The anti-alignment (alignment) of the total magnetic moment of Nd sublattice to the magnetic moment of Ni-Mn sublattice at low temperatures 
leads to $-\Delta S_{M} < (>) 0$ as seen in Fig. \ref{fig3}(a)-(c) (Fig. \ref{fig3}(d)-(f)) \cite{Das_2019}. In case of antialignment, external magnetic field can enforce an alignment and hence $\Delta S_{M}$ also shows a sign reversal at sufficiently large magnetic field (see, for example, Fig. \ref{fig3}(b)).
As a result, the IMCE at low temperatures changes to conventional MCE upon increasing the magnetic field strength. \par

From Fig. \ref{fig3} it can be seen that for  $R =$ Gd, Tb, Dy there exist two peaks with $-\Delta S_{M}>0$,  one at $T_{C}$ and other at low temperatures. Upon increasing magnetic field, both the peaks display characteristic features associated with convensional MCE, such as the broadening of temperature range and increase in the peak height. These observations are in accordance with the experimental findings.     

\section{Conclusions}
\label{Conclusion}
\noindent
In conclusion, we have presented a detailed experimental and theoretical study of the magnetization and magnetocaloric effect of rare-earth based double perovskites $R_2$NiMnO$_6$ ($R =$ Pr, Nd, Sm, Gd, Tb, and Dy) as a function of temperature and magnetic field. In contrast to Y$_2$NiMnO$_6$, which shows a single anomaly at the ferromagnetic transition temperature T$_{\rm C}$ caused by the ordering of Ni$^{2+}$ and Mn$^{4+}$ magnetic ions, all the $R_2$NiMnO$_6$ show anomalies in magnetization and MCE at T$_{\rm C}$, where the Ni-Mn sublattice orders ferromagnetically, as well as additional features in the magnetization and MCE at low temperature (T$_{\rm 2}$) because of the coupling of $R^{3+}$ ions and ordered Ni-Mn sublattice.   Pr$_2$NiMnO$_6$, Nd$_2$NiMnO$_6$ and Sm$_2$NiMnO$_6$ show a reduction in magnetization below T$_{\rm 2}$ which occurs inspite of the ferromagnetic coupling between the $R$ spins and the Ni/Mn spins.  This can be understood as the orbital moment is anti-aligned to the spin for the first half of the $R$ elements.  This decrease in magnetization below T$_{\rm 2}$ can be overcome by a large field strong enough to polarize the full effective moment of the $R^{3+}$ ions.  This leads to a negative value of (-$\Delta$S$_M$) in MCE measurements at  low temperature.  Just like the magnetization, the reduction of the MCE can be reversed on the application of a large enough magnetic field.  Materials from the second half of the $R$ series, $R_2$NiMnO$_6$ ($R =$ Gd, Tb, and Dy) show an upturn in the magnetization below T$_2$ because the orbital and spin moments of these $R$ ions are aligned with the Ni-Mn sublattice.  The large rare earth moments of for these heavier $R$ ions leads to a huge and positive value of (-$\Delta$S$_M$) at low temperature, which may potentially be useful in magnetic refrigeration.  Additionally, the paramagnetic background of rare-earth ions spreads out the magnetic entropy over a larger temperature range around T$_{\rm C}$ which broadens the MCE profile at T$_{\rm C}$ resulting in a large RCP.  The large RCPs make these materials potentially attractive for MCE based refrigeration techniques.  For the microscopic understanding of the magnetic behavior, we investigated a phenomenological two-sublattice Heisenberg model that takes into account both spin and orbital degrees of freedom. To study the model we use mean-field theory and classical Monte Carlo simulations. Both these methods reproduce the main features of the experimental observation for magnetization and for magnetic entropy change. We find that the magnetic behavior at low temperatures in the double perovskites containing f block elements is controlled by spin-orbit coupling. The more than or less than half-filled 4f orbitals decides the nature of spin-orbit coupling to be ferromagnetic or antiferromagnetic. The weak ferromagnetic coupling $J_{2}$ between the rare earth ion sublattice and Ni-Mn sublattice and antiferromagnetic spin-orbit coupling ($\lambda$) in DPs with $R =$ Nd, Pr and Sm is responsible for 
the downturn in magnetization at low temperature. On the other hand, the weak ferromagnetic coupling $J_{2}$ and ferromagnetic $\lambda$ in DPs with R = Gd, Tb and Dy explains the upturn in magnetization at low temperatures. 

\section{Acknowledgment}
\noindent
We acknowledge the support of the X-ray facility at IISER Mohali for powder XRD measurements. KP thanks DST, Government of India, for the award of Inspire faculty fellowship.

\bibliographystyle{apsrev4-1}
\bibliography{Ref}

\end{document}